\documentclass[12pt]{article}\usepackage{muxor-approx}
\begin{document}

\title{Quantum Compiling with  \\
Approximation of Multiplexors }

\author{Robert R. Tucci\\
        P.O. Box 226\\
        Bedford,  MA   01730\\
        tucci@ar-tiste.com}

\date{ \today}

\maketitle

\vskip2cm
\section*{Abstract}
A quantum compiling algorithm
is an algorithm for
decomposing (``compiling") an
arbitrary unitary matrix into
a sequence of elementary operations (SEO).
Suppose $U_{in}$
is an $\nb$-bit unstructured
unitary matrix
(a unitary matrix with no special symmetries)
that we wish to compile.
For $\nb>10$,
expressing $U_{in}$
as a SEO requires
more than a million CNOTs.
This calls for a
method for finding a
unitary
matrix that: (1)approximates $U_{in}$ well,
and (2 is
expressible with fewer CNOTs than $U_{in}$.
The purpose of this paper
is to propose
one such approximation method.
Various quantum compiling algorithms
have been proposed in the literature
that decompose
$U_{in}$ into
a sequence
of $U(2)$-multiplexors,
each of which is then decomposed
into
a SEO.
Our strategy for
approximating $U_{in}$ is
to approximate these intermediate
$U(2)$-multiplexors.
In this paper, we will show how
one can approximate
a $U(2)$-multiplexor
by another $U(2)$-multiplexor
that is expressible with fewer CNOTs.

 \newpage
\section{Introduction}

In quantum computing,
elementary operations are operations that
act on only a few (usually
one or two) qubits. For example, CNOTs and
one-qubit rotations are elementary operations.
A quantum compiling algorithm
is an algorithm for
decomposing (``compiling") an
arbitrary unitary matrix into
a sequence of elementary operations (SEO).
A quantum compiler is a software program
that implements a quantum compiling algorithm.

One measure of
the inefficiency of a quantum
compiler is
the number of CNOTs
it uses to express an unstructured
unitary matrix
(a unitary matrix with no special symmetries).
We will henceforth refer to
this number as $N_{CNOT}$.
Although good
quantum compilers
will also require optimizations that
deal with structured matrices,
unstructured matrices are certainly an
important case worthy of attention.
Minimizing the number of CNOTs
is a reasonable goal, since a
CNOT operation (or any 2-qubit
interaction used as a CNOT
surrogate) is
expected to take more time to perform
and to introduce more environmental
noise into the quantum computer
than a one-qubit rotation.
Ref.\cite{bound}
proved
that for matrices
of dimension $2^\nb$ ($\nb=$ number of bits),
$N_{CNOT}\geq \frac{1}{4}(4^\nb-3\nb-1)$.
This lower bound
is achieved for $\nb=2$
by the 3 CNOT circuits first
proposed in Ref.\cite{Vidal}.
It is not known whether
this bound can be achieved
for $\nb\geq 3$.
\begin{center}
{\scriptsize
\begin{tabular}{|r|r|r|}
\hline
$\nb$&
$\frac{1}{4}(4^\nb-3\nb-1)$& $4^{\nb-1}$ \\
\hline\hline
1 &  0.00 &   1 \\
2 &  2.25 &   4 \\
3 & 13.50  & 16 \\
4 &  60.75  & 64 \\
5  & 252.00 & 256 \\
6  & 1,019.25  &  1,024 \\
7 &  4,090.50  &  4,096 \\
8  & 16,377.75 &  16,384 \\
9 &  65,529.00 &  65,536 \\
10 & 262,136.25 & 262,144 \\
11 & 1,048,567.50  &  1,048,576 \\
12 & 4,194,294.75 &   4,194,304 \\
13  &16,777,206.00 &  16,777,216 \\
14 & 67,108,853.25 &  67,108,864 \\
15 & 268,435,444.50 & 268,435,456 \\
\hline
\end{tabular}
}
\end{center}

Suppose $U_{in}$
is an $\nb$-bit unstructured
unitary matrix that we wish to compile.
As the above table illustrates,
compiling $U_{in}$ is hopeless
for $\nb>10$ unless
we approximate
$U_{in}$.
We need a
method for finding a
unitary
matrix that: (1) approximates $U_{in}$ well,
and (2) is
expressible with fewer CNOTs than $U_{in}$.
The purpose of this paper
is to propose one such approximation
method.
The use of approximations
in quantum compiling
dates back to the earliest
papers in the field.
For example, Refs.\cite{Copper}
and \cite{Bar95}
contain discussions on this issue.
As Ref.\cite{Copper}
points out, even when compiling
a highly structured matrix
like the Discrete Fourier Transform
matrix,
some gates
that  contribute negligibly
to its exact SEO representation
can
be omitted with impunity.

Refs.\cite{Tuc99}\cite{Tuc04Oct}
and \cite{Tuc04Nov}
discuss a quantum compiling algorithm
that decomposes
an arbitrary unitary matrix into
a sequence
of $U(2)$-multiplexors,
each of which is then decomposed
into
a SEO.
Other workers
have proposed\cite{Mich04}\cite{Hels04b}
 alternative
quantum compiling
algorithms that
also generate
$U(2)$-multiplexors
as an intermediate
step.

\begin{figure}[h]
    \begin{center}
    \epsfig{file=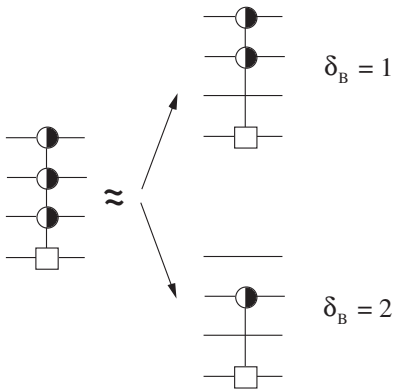, height=2.0in}
    \caption{
    Approximating a $U(2)$-multiplexor
    by another $U(2)$-multiplexor
    with $\delta_B$
    fewer controls.
    }
    \label{fig-approx}
    \end{center}
\end{figure}

The strategy proposed in this
paper for
approximating $U_{in}$ is
to approximate the intermediate
$U(2)$-multiplexors
whose product equals $U_{in}$.
In this paper, we will show how
one can approximate
a $U(2)$-multiplexor
by another $U(2)$-multiplexor
(the ``approximant")
that has fewer controls, and, therefore,
is expressible with fewer CNOTs.
We
will
call the
 reduction in the number
of control bits the
{\bf bit deficit} $\delta_B$.
Fig.\ref{fig-approx}
is emblematic of
our approach.
It shows a $U(2)$-multiplexor
with 3 controls being
approximated by
either a
$U(2)$-multiplexor
with 2 controls or one
with 1 control.

\section{Notation}\label{sec-notation}
In this section, we will
define some notation that is
used throughout this paper.
For additional information about our
notation, we recommend that
the reader
consult Ref.\cite{Paulinesia}.
Ref.\cite{Paulinesia} is
a review article, written
by the author of this paper, which
uses the same notation as this paper.

Let $Bool =\{0, 1\}$.
As usual, let $\ZZ, \RR, \CC$ represent the set
of integers (negative  and non-negative),
real numbers, and
complex numbers, respectively.
For integers $a$, $b$
 such that $a\leq b$, let
$\ZZ_{a,b}=\{a, a+1,
\ldots b-1, b\}$.
For $\Gamma$ equal to $\ZZ$ or $\RR$, let
$\Gamma^{>0}$ and  $\Gamma^{\geq 0}$
represent the set of
positive
and
non-negative $\Gamma$
numbers, respectively.
For any positive integer $n$
and any set $S$, let
$S^n$ denote
the Cartesian product of
$n$ copies of $S$; i.e.,  the set of
all $n$-tuples
of elements of $S$.

For any (not necessarily distinct)
objects $a_1, a_2, a_3, \ldots$, let
$\{a_1, a_2, a_3,\ldots\}_{ord}$
denote an ordered set.
For some object $b$, let
$b\{a_1, a_2, a_3,\ldots\}_{ord}=
\{ba_1, ba_2, ba_3,\ldots\}_{ord}$.
Let $\emptyset$ be the empty set.
For an ordered set $S$,
let $S^R$ be $S$ in reverse order.

We will use $\theta(S)$
to represent the ``truth function";
$\theta(S)$ equals 1 if statement $S$ is true
and 0 if $S$ is false.
For example, the Kronecker delta
function is defined by
$\delta^y_x=\delta(x,y) = \theta(x=y)$.
For $x\in Bool$,
\beq
\sum_{k=0}^{1}(-1)^{kx}=2
\delta(x,0)
\;.
\label{eq-bool-delta-fun}
\eeq

For any positive integer $N$,
we will use $\vec{e}_i$ where
$i=1, 2, \ldots, N$ to
denote the standard basis vectors
in $N$ dimensions; i.e.,
$[\vec{e}_i]_j = \delta(i,j)$
for $i,j\in \ZZ_{1, N}$.

$I_n$ and $0_n$ will represent the
$n$-dimensional unit and zero matrices.

For any matrix $A$ and positive integer $n$,
let

\beq
A^{\otimes n} =
\underbrace{A\otimes \cdots
\otimes A \otimes A}_{n\;\;{\rm copies\;\;of\;\;} A}
\;,
\eeq

\beq
A^{\oplus n} =
\underbrace{A\oplus \cdots
\oplus A \oplus A}_{n\;\;{\rm copies\;\;of\;\;} A}
\;.
\eeq

For any matrix $A\in \CC^{m\times n}$
and $p=1,2,\infty$,
$\|A\|_p$
will represent the $p$-norm of $A$, and
$\|A\|_F$ its Frobenius
norm. See \cite{Golub} for
a discussion of matrix norms.

Let
$\vec{x}\in \CC^{n\times1}$.
As is customary in
the Physics literature,
$\|\vec{x}\|_2$
will also be denoted
by $|\vec{x}|$
and called the magnitude of $\vec{x}$.
For any complex matrix $A$,
we will use $abs(A)$ to denote
the matrix that is obtained from
$A$ by replacing each of its
entries
by its absolute value. In other words,
$[abs(A)]_{ij}= abs(A_{ij}) = |A_{ij}|$.
(Careful: Ref.\cite{Golub} and many
other mathematical books call $|A|$
what we call $abs(A)$).

Any $x\in \RR$ can be
expressed as a doubly infinite
power series in powers of
a  base $E\in \ZZ^{>0}$:
$x=\pm\sum_{\alpha =-\infty}^\infty
x_\alpha E^\alpha$.
This expansion can be represented
by:
$\pm (\cdots x_1 x_0.
x_{-1} x_{-2}\cdots)_{\flat E}$,
which is called
the  base $E$ representation of $x$.
The plus or minus in
these expressions is chosen
to agree with the sign of $x$.
It is customary to omit
the subscript $\flat E$
when $E=10$. For example
$2.25 = 2 +\frac{1}{4} = \bin{1.01}$

Suppose $x=\rbin{x}
\in \RR^{\geq 0}$.
Note that division by 2
shifts the binary representation
of $x$ one space to the right:
$\frac{x}{2} = \bin{\cdots x_1.
x_0 \cdots}$.
Likewise, multiplication by
2 shifts the binary representation of $x$
one space to the left:
$2x= \bin{\cdots x_{-1}.x_{-2} \cdots}$.
In general, for any $\alpha\in \ZZ$,
$\frac{x}{2^\alpha} =
\bin{\cdots x_\alpha.x_{\alpha-1}
 \cdots}$.

Define the action
of an overline placed
over an $a\in Bool$
by
$\overline{0}=1$ $\overline{1}=0$.
Call this bit negation.
Define the action
of an oplus placed between
$a,b\in Bool$ by
$a\oplus b = \theta(a\neq b)$.
Call this bit addition.
One can extend the
bit negation and
bit addition
operations so that
they can act on non-negative reals.
Suppose $x=\rbin{x}$,
and  $y=\rbin{y}$ are non-negative real numbers.
Then define the action
of an overline over $x$
so that it acts on each bit individually; i.e.,
so that
$[\overline{x}]_\alpha = \overline{x_\alpha}$.
This overline operation is sometimes
called bitwise negation.
Likewise, define the action of an
oplus placed between $x$ and $y$ by
$(x\oplus y)_\alpha=
x_\alpha\oplus y_\alpha$.
This oplus operation is sometimes
called bitwise addition
(without carry).

For any $x\in \RR$, the floor function
is defined by
$\floor{x} =
\max \{j\in \ZZ : j\leq x\}$,
and the ceiling function
by
$\ceil{x} =
\min \{j\in \ZZ : j\geq x\}$.
For example, if $x=\rbin{x}$,
then $\floor{x} = \ibin{x}$.

We will often use $\nb$ to denote a
number of bits, and $\ns=2^{\nb}$
to denote
the corresponding number of states.
We will use the sets
$Bool^\nb$ and $\ZZ_{0, \ns-1}$
interchangeably,
since any $x\in \ZZ_{0, \ns-1}$
can be identified with
its binary representation
$\bin{x_{\nb-1}\cdots x_1 x_0}\in Bool^{\nb}$.

For any $x=
\bin{x_{\nb-1}\cdots x_1 x_0}
\in \ZZ_{0, \ns-1}$,
define $x^R=\bin{x_0 x_1 \cdots x_{\nb-1}}$;
i.e., $x^R$ is the result of
reversing the binary representation of $x$.

Suppose
$\pi:\ZZ_{0, \ns-1}\rarrow \ZZ_{0, \ns-1}$
is a 1-1 onto map.
(We use the letter $\pi$
to remind us it is a permutation;
i.e., a 1-1 onto map from a
finite set onto itself).
One can define a
permutation matrix $M$
with entries
given by
$M_{yx} = \theta(y=\pi(x))$
for all $x,y\in \ZZ_{0, \ns-1}$.
(Recall that all permutation matrices $M$
arise from permuting the columns of
the unit matrix, and they satisfy $M^TM=1$.)
In this paper,
we will often represent
the map $\pi$
and its corresponding
matrix $M$ by the same symbol $\pi$.
Whether the function  or
the matrix is being
alluded to will be clear from the
context.
For example, suppose $A$ is an $\ns$
dimension matrix, and $\pi$
 is a permutation on the set $\ZZ_{0, \ns-1}$.
Then, it is easy to check
that for all $i,j\in \ZZ_{0, \ns-1}$,
$(\pi^T A)_{ij} = A_{\pi(i),j}$ and
$(A\pi)_{ij} = A_{i,\pi(j)}$.

Suppose $\pi_B:\ZZ_{0,\nb-1}\rarrow
\ZZ_{0,\nb-1}$ is a 1-1 onto map
(i.e., a bit permutation).
$\pi_B$ can be extended to a map
$\pi_B:\ZZ_{0,\ns-1}\rarrow
\ZZ_{0,\ns-1}$
as follows. If
$x=\bin{x_{\nb-1}
\cdots x_1 x_0}\in \ZZ_{0,\ns-1}$,
then let
$[\pi_B(x)]_\alpha = x_{\pi_B(\alpha)}$
for all $\alpha\in \ZZ_{0,\nb-1}$.
The function
$\pi_B:\ZZ_{0,\ns-1}\rarrow
\ZZ_{0,\ns-1}$ is 1-1 onto,
so it can be used
to define a permutation matrix
of the same name.
Thus, the symbol
$\pi_B$ will be used
to refer to 3 different objects:
a permutation on the set $\ZZ_{0, \nb-1}$,
a permutation on the set $\ZZ_{0, \ns-1}$,
and an $\ns$ dimensional permutation matrix.
All permutations on
$\ZZ_{0, \nb-1}$ generate
a permutation on
$\ZZ_{0, \ns-1}$, but not all
permutations on
$\ZZ_{0, \ns-1}$
have an underlying
permutation on
$\ZZ_{0, \nb-1}$.

An example of a
bit permutation that will
arise later is
$\pi_R$; it
maps
$\pi_R(i)=i^R$ for
all $i\in \ZZ_{0,\ns-1}$
and
$\pi_R(\alpha)=\nb-1-\alpha$
for all $\alpha\in \ZZ_{0, \nb-1}$.

\section{Gray Code}

In this section, we will
review some well known facts about
Gray code\cite{Knuth4}.
(Gray code was named after a person
named Gray, not after the color.)

For any positive integer $\nb$,
we define a {\bf Grayish code} to be
a list of the elements
of $Bool^\nb$ such that adjacent
$\nb$-tuples of the
list differ in only  one component.
In other words, a Grayish code
is a 1-1 onto map
$\pi_{Gish}:\ZZ_{0, \ns-1}\rarrow
\ZZ_{0, \ns-1}$
such that,
for all $k\in \ZZ_{0, \ns-2}$,
the binary representations of
$\pi_{Gish}(k)$ and $\pi_{Gish}(k+1)$
differ in only one component.
For any $\nb>1$, there are many
functions $\pi_{Gish}$ that satisfy this
definition.

Next we
will define a particular
Grayish code that we
shall refer to as ``the" Gray code
and denote by $\pi_G$.
The Gray
code for $\nb=1, 2, 3$ is:

\beq
\begin{array}{|c|c|c|}
\hline
k&\bin{k}&\pi_G(k)\\
\hline
0&0&0\\
1&1&1
\\ \hline
\end{array}
\;\;
\begin{array}{|c|c|c|}
\hline
k&\bin{k}&\pi_G(k)\\
\hline
0&00&00\\
1&01&01\\
2&10&11\\
3&11&10
\\ \hline
\end{array}
\;\;
\begin{array}{|c|c|c|}
\hline
k&\bin{k}&\pi_G(k)\\
\hline
0&000&000\\
1&001&001\\
2&010&011\\
3&011&010\\
4&100&110\\
5&101&111\\
6&110&101\\
7&111&100
\\ \hline
\end{array}
\;.
\eeq
The {\bf Gray code} can be defined
recursively as follows.
Let $\Gamma_0 =\emptyset$. For $\nb>0$,
let $\Gamma_\nb$ equal the set $Bool^\nb$
ordered in the Gray code order.
In other words,
$\Gamma_\nb = \{\pi_G(0), \pi_G(1),
\pi_G(2), \ldots , \pi_G(2^\nb-1)\}_{ord}$.
 Then,

\beq
\Gamma_{\nb+1} =
\{0\Gamma_\nb, 1\Gamma^R_\nb\}_{ord}\
\;
\eeq
for $\nb\in\ZZ_{0, \infty}$.(See Section
\ref{sec-notation} for ordered set notation.)

From the
recursive definition of the Gray
code, it is possible to
prove that
if $k=\ibin{k}$ and $g=\ibin{g}$
are
nonnegative integers such
that $g=\pi_G(k)$,
then

\begin{subequations}
\label{eq-gk-comps}
\beq
g_\alpha = k_\alpha
\oplus k_{\alpha+1}
\;,
\label{eq-g-in-k-comps}
\eeq
for all $\alpha\in \ZZ_{0, \infty}$.
(For  all $\alpha>\nb-1$,
$k_\alpha=g_\alpha=0$).
Eq.(\ref{eq-g-in-k-comps})
specifies $\nb$ linear
equations for the
$\nb$ components of $g$
expressed in terms of
the $\nb$ components of $k$.
These equations can be easily
inverted using Gauss Elimination
to get:

\beq
k_\alpha = g_\alpha
\oplus g_{\alpha+1}
\oplus g_{\alpha+2}
\oplus g_{\alpha+3}
\oplus \cdots
\;.
\label{eq-k-in-g-comps}
\eeq
\end{subequations}
Eqs.(\ref{eq-gk-comps})
can also be written in terms of  the
floor function:

\begin{subequations}
\label{eq-gk-floor}
\beq
g = k
\oplus \floor{\frac{k}{2}}
\;,
\eeq

\beq
k = g
\oplus \floor{\frac{g}{2}}
\oplus \floor{\frac{g}{2^2}}
\oplus \floor{\frac{g}{2^3}}
\oplus \cdots
\;.
\eeq
\end{subequations}

As in Section \ref{sec-notation},
suppose $\pi_B$ represents
a permutation on
$\ZZ_{0, \nb-1}$
which generates a
permutation on
$\ZZ_{0, \ns-1}$
of the same name.
Clearly, $\pi_B\circ\pi_G$
is a Grayish code. Indeed,
$\pi_B\circ\pi_G$ is a 1-1 onto map,
and permuting bits
the same way for all elements of a list
preserves the property that
adjacent $\nb$-tuples
differ in only one component.
(Note, however,  that it is easy
to find $\pi_B$'s  such that $\pi_G\circ\pi_B$
is not a Grayish code.
Hence, to preserve Grayishness,
one must apply the bit permutation
after $\pi_G$, not before).

\section{Hadamard, Paley and Walsh Matrices}
In this section, we will review some
well known facts about the so called
Hadamard, Paley and Walsh matrices
(a.k.a. transforms) \cite{Knuth4}.

For any positive integer
$\nb$,
we define
the $\nb$-bit Hadamard matrix by

\beq
(H_\nb)_{k,r}=
\frac{1}{\sqrt{\ns}}
(-1)^{
\sum_{\alpha=0}^{\nb-1}
k_\alpha r_\alpha
},
\;
\eeq
the $\nb$-bit Paley matrix by

\beq
(P_\nb)_{k,r}=
\frac{1}{\sqrt{\ns}}
(-1)^{
\sum_{\alpha=0}^{\nb-1}
\sum_{\beta=0}^{\nb-1}
k_\alpha r_\beta
\delta_{\alpha + \beta}^{\nb-1}
},
\;
\eeq
and the $\nb$-bit Walsh matrix by

\beq
(W_\nb)_{k,r}=
\frac{1}{\sqrt{\ns}}
(-1)^{
\sum_{\alpha=0}^{\nb-1}
\sum_{\beta=0}^{\nb-1}
k_\alpha r_\beta
[
\delta_{\alpha + \beta}^{\nb}
+
\delta_{\alpha + \beta}^{\nb-1}]
}
\;,
\eeq
where
$k,r\in \ZZ_{0, 2^\nb-1}$,
$k=\ibin{k}$ and $r=\ibin{r}$.
We will often omit
the subscript $\nb$
from $H_\nb, P_\nb, W_\nb$
in contexts  where
doing this does not lead to confusion.

Note that $H, P, W$ are
real symmetric matrices.

For $j\in \ZZ_{0,\ns-1}$, define
the ``reversal" function
$\pi_{R(\nb)}(j)=j^R$,
and the ``negation" function
$\pi_{N(\nb)}(j)=\overline{j}$.
The function $\pi_{G(\nb)}$
for $\nb$-bit Gray code
has been defined
previously.
The functions
$\pi_{R(\nb)}$,
$\pi_{N(\nb)}$
and
$\pi_{G(\nb)}$
are 1-1 onto so they
can be used to define
permutation matrices
of the same name (See Section
\ref{sec-notation}.)
We will often write
$\pi_R$,
$\pi_N$
and
$\pi_G$
instead of
$\pi_{R(\nb)}$,
$\pi_{N(\nb)}$
and
$\pi_{G(\nb)}$
in contexts where this
does not lead to confusion.

Note that
$\pi_R$ and
$\pi_N$
are symmetric matrices
but
$\pi_G$ isn't.

Next we will show that the $\nb$-bit
Hadamard, Paley and Walsh matrices
all have the same columns,
except in different orders.
More specifically,
$H,P, W$
are related to each other
by the following equations:

\beq
H\pi_R=P
\;,\;\;
P\pi_G =W
\;.
\label{eq-h-p-w-relations}
\eeq
A more pictorial way
of expressing Eqs.(\ref{eq-h-p-w-relations})
is:

\beq
H
\stackrel{(\cdot)\pi_R}{\longrightarrow}
P
\stackrel{(\cdot)\pi_G}{\longrightarrow}
W
\;.
\eeq
Taking the transpose
of both sides of Eqs.(\ref{eq-h-p-w-relations})
leads to

\beq
\pi_R H =P
\;,\;\;
\pi_G^T  P=W
\;.
\label{eq-h-p-w-transp-relations}
\eeq
In the last equation,
we have used the fact
that matrices
$H,P,W$,
$\pi_R, \pi_N$
are symmetric
but $\pi_G$ isn't.

Comparing Eqs.(\ref{eq-h-p-w-relations})
and (\ref{eq-h-p-w-transp-relations}),
we see that

\beq
H\pi_R = \pi_R H
\;,
\label{eq-hs-r-sym}
\eeq
and

\beq
P\pi_G = \pi_G^T P
\;.
\label{eq-ps-g-sym}
\eeq
In fact, Eq.(\ref{eq-hs-r-sym})
can be generalized as
follows. Suppose $\pi_B$ is
a bit permutation on $\ZZ_{0, \nb-1}$.
Then

\beqa
(\pi_B^T H \pi_B)_{kr}
&=&
H_{\pi_B(k),\pi_B(r)}\\
&=&
\frac{1}{\sqrt{\ns}}
(-1)^{
\sum_{\alpha=0}^{\nb-1}
[\pi_B(k)]_\alpha
[\pi_B(r)]_\alpha
}\\
&=&
\frac{1}{\sqrt{\ns}}
(-1)^{
\sum_{\alpha=0}^{\nb-1}
k_{\pi(\alpha)}
r_{\pi(\alpha)}
}\\
&=&
H_{kr}
\;,
\eeqa
so

\beq
H\pi_B = \pi_B H
\;.
\label{eq-hs-b-sym}
\eeq
Eq.(\ref{eq-hs-b-sym}) becomes
Eq.(\ref{eq-hs-r-sym})
when $\pi_B = \pi_R$.

To prove Eqs.(\ref{eq-h-p-w-relations}),
note that

\beqa
(H\pi_R)_{ik}&=&
\sum_{j=0}^{\ns-1} H_{ij}(\pi_R)_{jk}\\
&=&
\frac{1}{\sqrt{\ns}}
\sum_j(-1)^{\sum_\alpha
i_\alpha j_\alpha}
\theta(j=k^R)\\
&=&P_{ik}
\;.
\eeqa
Similarly,

\beqa
(P\pi_G)_{ik}&=&
\sum_{j=0}^{\ns-1} P_{ij}(\pi_G)_{jk}\\
&=&
\frac{1}{\sqrt{\ns}}
\sum_j(-1)^{\sum_{\alpha,\beta}
i_\alpha j_\beta
\delta_{\alpha+\beta}^{\nb-1}}
\theta(j=\pi_G(k))\\
&=&
\frac{1}{\sqrt{\ns}}
(-1)^{\sum_{\alpha,\beta}
i_\alpha
(k_\beta \oplus k_{\beta+1})
\delta_{\alpha+\beta}^{\nb-1}}
\\
&=&
\frac{1}{\sqrt{\ns}}
(-1)^{
\sum_{\alpha}
i_\alpha
(k_{\nb-1-\alpha} + k_{\nb-\alpha})}     \\
&=&
W_{ik}
\;.
\eeqa

The square of $H, P$ and $W$
is one. Indeed,
using Eq.(\ref{eq-bool-delta-fun})
we get

\beqa
(H^2)_{ik}&=&
\sum_{j=0}^{\ns-1} H_{ij}H_{jk}\\
&=&
\frac{1}{2^{\nb}}
\sum_{j_{\nb-1}=0}^1\ldots
\sum_{j_1=0}^1
\sum_{j_0=0}^1
(-1)^{
\sum_{\alpha=0}^{\nb-1}
(i_\alpha+k_\alpha)j_\alpha}\\
&=&\prod_\alpha \delta_{i_\alpha}^{k_\alpha}
=\delta_{i}^{k}
\;,
\eeqa

\beq
P^2=(\pi_R H)(H\pi_R)=1
\;,
\eeq
and

\beq
W^2=(\pi_G^T P)(P\pi_G)=1
\;.
\eeq
Since their square equals one,
and they are real symmetric matrices,
$H, P$ and $W$ are also
orthogonal matrices.

From the
definitions given above for
$\pi_R, \pi_N,
\pi_G$, one can prove by induction
on $\nb$ that
these matrices obey the following
recursive equations:

\begin{subequations}
\beq
\pi_{R(0)}=1
\;,
\;\;
\pi_{R(\nb+1)}=
\left[
\begin{array}{c}
\pi_{R(\nb)}\otimes(1,0)\\
\pi_{R(\nb)}\otimes(0,1)
\end{array}
\right]
\;,
\eeq

\beq
\pi_{N(0)}=1
\;,
\;\;
\pi_{N(\nb+1)}=
\pi_{N(\nb)}
\otimes
\left[
\begin{array}{cc}
0&1\\
1&0
\end{array}
\right]
\;,
\eeq
and

\beq
\pi_{G(0)}=1
\;,
\;\;
\pi_{G(\nb+1)}=
\left[
\begin{array}{cc}
\pi_{G(\nb)}&0\\
0&\pi_{G(\nb)}\pi_{N(\nb)}
\end{array}
\right]
\;.
\eeq
\end{subequations}
Similarly,
from the
definitions given above for
$H, P, W$, one can prove by
induction on $\nb$ that
these matrices obey the following
recursive equations:

\begin{subequations}
\beq
H_{0} = 1
\;,\;\;
H_{\nb+1} =
H_{\nb}\otimes
\left[
\begin{array}{cc}
1&1\\
1&-1
\end{array}
\right]
\frac{1}{\sqrt{2}}
\;,
\eeq

\beq
P_{0} = 1
\;,\;\;
P_{\nb+1}=
\left[
\begin{array}{cc}
P_{\nb}\otimes(1,1)\\
P_{\nb}\otimes(1,-1)
\end{array}
\right]
\frac{1}{\sqrt{2}}
\;,
\eeq
and

\beq
W_{0} = 1
\;,\;\;
W_{\nb+1}=
\left[
\begin{array}{cc}
W_{\nb}\otimes(1,1)\\
(\pi_{N(\nb)}W_{\nb})\otimes(1,-1)
\end{array}
\right]
\frac{1}{\sqrt{2}}
\;.
\eeq
\end{subequations}

By virtue of Eqs.(\ref{eq-h-p-w-relations}),

\beq
W = H \pi_R \pi_G
\;,
\label{eq-w-fun-h}
\eeq
Eq.(\ref{eq-w-fun-h}) means that the permutation
$\pi_R \pi_G$ will permute
the columns of $H$ to give $W$.
Expressing Eq.(\ref{eq-w-fun-h})
in component
form, we find

\beqa
W_{ij} &=&
\sum_{r,k} H_{ir}
(\pi_R)_{rk}
(\pi_G)_{kj}\\
&=&
\sum_{r,k} H_{ir}
\theta(r=\pi_R(k))
\theta(k=\pi_G(j))\\
&=& H_{i,\pi_R\circ \pi_G(j)}
\;.
\eeqa
Thus, if we denote the
columns of $H_\nb$ and $W_\nb$
by $\vec{h}_j$ and $\vec{w}_j$,
respectively, then

\beq
\vec{w}_j = \vec{h}_{\pi_R\circ \pi_G(j)}
\;,
\label{eq-wi-def}
\eeq
for $j\in \ZZ_{0, \ns-1}$.

\section{Constancy}
In this section, we will define
a property of vectors
called constancy.
The columns of $H_\nb$
can be conveniently classified according
to their constancy.

Consider the 3-bit
Hadamard matrix:

\beq
H_3 = H_1^{\otimes 3}=
\frac{1}{\sqrt{2^3}}
\begin{array}{|rrrrrrrr|}
{\scriptstyle\vec{h}_{000}}&
{\scriptstyle\vec{h}_{001}}&
{\scriptstyle\vec{h}_{010}}&
{\scriptstyle\vec{h}_{011}}&
{\scriptstyle\vec{h}_{100}}&
{\scriptstyle\vec{h}_{101}}&
{\scriptstyle\vec{h}_{110}}&
{\scriptstyle\vec{h}_{111}}
\\
\hline
\pone&\pone&\pone&\pone&\pone&\pone&\pone&\pone
\\
\pone&\none&\pone&\none&\pone&\none&\pone&\none
\\
\pone&\pone&\none&\none&\pone&\pone&\none&\none
\\
\pone&\none&\none&\pone&\pone&\none&\none&\pone
\\
\pone&\pone&\pone&\pone&\none&\none&\none&\none
\\
\pone&\none&\pone&\none&\none&\pone&\none&\pone
\\
\pone&\pone&\none&\none&\none&\none&\pone&\pone
\\
\pone&\none&\none&\pone&\none&\pone&\pone&\none\\
\hline
\end{array}
\;,
\label{eq-had3}
\eeq
where we have labelled the columns of $H_3$
by $\vec{h}_j$,
where the index $j$ is given
in its binary representation.
According to Eq.(\ref{eq-wi-def}),
to  get $W_3$ from Eq.(\ref{eq-had3}),
one can simply reorder
the columns of $H_3$ in bit-reversed Gray code.
The columns of $H_3$ (and of $W_3$)
can be classified according
to their constancy.
We define the
{\bf constancy} $\calc (\vec{h})$
of a vector $\vec{h}$ to be
the smallest number of
identical adjacent
entries of $\vec{h}$. For
example,
$\calc ([1,-1,1,1]^T)=1$
and
$\calc ([1,1,-1,-1]^T)=2$.
The next table gives
the constancy of the columns
of $H_3$, with the columns
listed
in the order in which they
appear in $W_3$.

\beq
\begin{array}{|l|l|l|l|}
\hline
k&\pi_G(k)&\pi_R\circ \pi_G(k)
&\calc[\vec{h}_{\pi_R\circ \pi_G(k)}]\\
\hline
\hline
000&000&000&8\\  \hline
001&001&100&4\\  \hline
010&011&110&2\\  \hline
011&010&010&2\\  \hline
100&110&011&1\\  \hline
101&111&111&1\\  \hline
110&101&101&1\\  \hline
111&100&001&1\\
\hline
\end{array}
\;.
\label{tab-const}
\eeq
It is clear from Eq.(\ref{tab-const})
that the columns of $W_3$
are listed in order of
non-increasing constancy,
and that the constancies of
the columns of
$W_3$ are all powers of 2.
The literature
on Walsh matrices often refers to
the index that labels the columns
of $W$ as the {\bf sequency} of that column.
Thus, as sequency increases,
constancy decreases or stays the same.
Sequency and Constancy are
analogous to Frequency and Period,
respectively, in Fourier Analysis.

Note that given any
matrix $A$,
more than one of the columns of $A$
may have the same constancy.
We will refer to:
the number of columns of $A$
with the same constancy $K$, as:
the {\bf multiplicity}
{\bf of the
constancy} $K$ {\bf in the matrix} $A$,
and denote it by $\calm_A(\calc=K)$.
In this paper, we are only
 concerned with
the case where $A$ equals the
$\nb$-bit Hadamard matrix so
we will henceforth omit the
subscript $A$ from
 $\calm_A(\calc=K)$. Sometimes we
 will also abbreviate $\calm(\calc=K)$
 by $\calm(K)$, if doing this does not
 lead to
 confusion. The next table
 gives the multiplicity of the constancy $K$
 in the $\nb$-bit Hadamard matrix:

\beq
\begin{array}{l|lllll}
\calm(\calc=K)&K=1 & K=2& K=4& K=8&\cdots\\
\hline
\nb=1 & 1 & 1 & 0 & 0&\cdots\\
\nb=2 & 2 & 1 & 1 & 0&\cdots\\
\nb=3 & 4 & 2 & 1 & 1&\cdots\\
\vdots & \vdots& \vdots
& \vdots & \vdots & \vdots \\
\end{array}
\;.
\label{tab-multi}
\eeq
It is also convenient to define

\beq
\calm(\calc\geq \ns')=
\sum_{K\geq \ns'}
\calm(K)
\;.
\label{eq-cum-multi}
\eeq
We shall call this the
{\bf cumulative multiplicity of the constancy}.
The next table
can be easily obtained from
Eq.(\ref{tab-multi}) and Eq.(\ref{eq-cum-multi}).
It
gives
$\calm(\calc\geq \ns')$ for the
$\nb$-bit Hadamard matrix.

\beq
\begin{array}{l|lllll}
\calm(\calc\geq \ns')&\ns'=2^0 & \ns'=2^1& \ns'=2^2& \ns'=2^3&\cdots\\
\hline
\nb=1 & 2 & 1 & 0 & 0&\cdots\\
\nb=2 & 4 & 2 & 1 & 0&\cdots\\
\nb=3 & 8 & 4 & 2 & 1&\cdots\\
\vdots & \vdots& \vdots
& \vdots & \vdots & \vdots \\
\end{array}
\;.
\label{tab-cum-multi}
\eeq
It is clear from Eq.(\ref{tab-cum-multi})
that $\calm(\calc\geq \ns') =
2^{\nb-\nb'}\theta(\nb\geq \nb')$,
where $\ns'=2^{\nb'}$.

\section{Symmetries of Multiplexors}
In this section,
we
discuss some symmetries of exact
decompositions of
$U(2)$-multiplexors.

For simplicity, we will first
consider the $R_y(2)$-multiplexors
used in Ref.\cite{Tuc99}.
Ref.\cite{Tuc04Oct} uses
$U(2)$-multiplexors that are
more general than the
$R_y(2)$-multiplexors
used in Ref.\cite{Tuc99}.
At the end of the
paper, we will discuss
how to generalize our
results for $R_y(2)$-multiplexors
so that they apply to
 the more general multiplexors used
in Ref.\cite{Tuc04Oct}.

Below, we
will present some
quantum circuit diagrams.
Besides the
circuit notational conventions defined
in Refs.\cite{Tuc04Oct} and
\cite{Paulinesia},
the circuits below will
use the following additional notation.
A square gate with an angle $\theta$
below the square will
represent
$\exp(i\theta\sigy)$
applied at that ``wire".
Typically, we will consider a
SEO
consisting of alternating
one-qubit rotations
and CNOTs. The SEO will
always have a one-qubit
rotation at one end and a CNOT
at the other. The
angle for the one-qubit rotation
that either begins or ends the SEO
will be denoted by $\theta_{00\ldots0}$.
Given
two adjacent angles $\theta_b$
and $\theta_{b'}$ in the SEO,
$\bin{b}$
and $\bin{b'}$
will differ only
in one
component,
component $\alpha$,
where $\alpha$ is the
position
of
the control bit
of the CNOT that
lies between
the $\theta_b$ and
$\theta_{b'}$ gates.

If we take
the Hermitian  conjugate
of
the multiplexor
$\exp(i\sum_{b\in Bool^{\nb-1}} \phi_{b}
 \sigy\otimes P_{b})$,
and then we replace
the angles $\phi_{b}$
by their negatives
(and also the angles $\theta_b$, Hadamard
transforms of the $\phi_b$, by their negatives),
we get the same multiplexor back.
Henceforth,
we will refer to this
symmetry transformation
as {\bf time reversal}.
Thus,
an $R_y(2)$-multiplexor
is
invariant under
time reversal.

Suppose $\pi_B$
is a bit permutation on $\nb-1$ bits.
If we replace
$\phi_{b}$ by
$\phi_{\pi_B(b)}$
(and also $\theta_b$ by
$\theta_{\pi_B(b)}$)
and
$P_{b}$ by
$P_{\pi_B(b)}$
in
the multiplexor
$\exp(i\sum_{b\in Bool^{\nb-1}} \phi_{b}
 \sigy\otimes P_{b})$,
we get the same multiplexor back.
Henceforth,
we will refer to this
symmetry transformation
as {\bf bit permutation}.
Thus,
an $R_y(2)$-multiplexor
is
also invariant under
bit permutation.

\begin{figure}[h]
    \begin{center}
    \epsfig{file=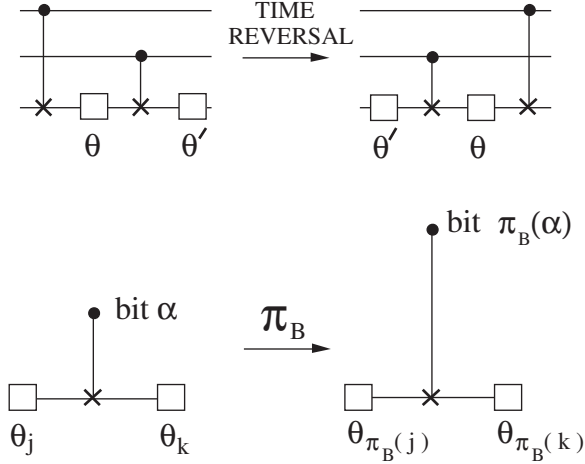, height=2.5in}
    \caption{
    Examples of
    the action of
    time
    reversal and
    bit permutation
    on a string of one-qubit
    rotations and CNOTs.
    }
    \label{fig-sym-ops}
    \end{center}
\end{figure}

Fig.\ref{fig-sym-ops}
shows how time reversal and bit permutation
act on a sequence of
one-qubit rotations and CNOTs.
More examples of the
application of these transformations
will be given below.

\begin{figure}[h]
    \begin{center}
    \epsfig{file=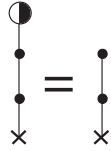, height=.75in}
    \caption{
    A half-moon node
    represents a projector $P_b$
    where $b\in Bool$.
    The half-moon node may be omitted
    when it appears in a multiplexor
    whose $U(2^{N_T})$-subset does not depend
    on the index $b$. This figure is an
    example of this principle.
    }
    \label{fig-free-half-moon}
    \end{center}
\end{figure}

Recall from Ref.\cite{Tuc04Oct}
our definition of a general
multiplexor
with $N_K$ control qubits $\vec{\kappa}$
and $N_T$ target qubits $\vec{\tau}$:
$\sum_{{b}\in Bool^{N_K}}
U_{b}(\vec{\tau})
P_{b}(\vec{\kappa})$.
In a multiplexor whose matrices
$U_{b}$
are independent of
the $\alpha$ component
$b_\alpha$ of $b$,
we can sum $P_{b_\alpha}$
over $b_\alpha\in Bool$
to get 1. Such a multiplexor
acts as the identity on
qubit $\alpha$.
When representing such a multiplexor
in a circuit diagram,
we can omit its half-moon
node on qubit line $\alpha$.
Fig.\ref{fig-free-half-moon}
shows a very special case of
this principle,
a special case that will
be used in the circuit diagrams
below.

\begin{figure}[h]
    \begin{center}
    \epsfig{file=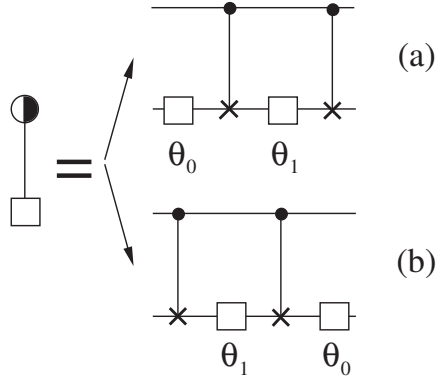, height=2.0in}
    \caption{
    Two possible decompositions
    of an $R_y(2)$-multiplexor
    with 1 control.
    }
    \label{fig-1control}
    \end{center}
\end{figure}

Fig.\ref{fig-1control}
shows two possible ways of
decomposing an $R_y(2)$-multiplexor
with one control.
The
decomposition (a) in
Fig.\ref{fig-1control}
is equivalent to:

\beq
\exp\left(i\sum_{b\in Bool}
\phi_b\sigy\otimes P_b\right)=
e^{i\theta_0\sigy(1)}
\sigx(1)^{n(0)}
e^{i\theta_1\sigy(1)}
\sigx(1)^{n(0)}
\;.
\label{eq-decomp-1control}
\eeq
Let LHS and RHS
stand for the left and right hand
sides of Eq.(\ref{eq-decomp-1control}).
Recall that $n=diag(0,1)=P_1$
and $\nbar= 1-n = diag(1,0)=P_0$.
Eq.(\ref{eq-decomp-1control})
can be proven as follows:

\begin{subequations}
\label{eq-decomp-1control-proof}
\begin{eqnarray}
RHS
&=&
e^{i\theta_0\sigy(1)}
e^{i\theta_1\sigy(1)\sigz(0)}\\
&=&
e^{i\sigy(1)\{\theta_0 +
\theta_1 [P_0(0)-P_1(0)]\}}\\
&=&LHS \label{eq-step-subst-theta}
\;.
\end{eqnarray}
\end{subequations}
To arrive at Eq.(\ref{eq-step-subst-theta}),
we
expressed $\theta_0, \theta_1$
in terms of $\phi_0, \phi_1$
using

\beq
\left[
\begin{array}{c}
\theta_0\\
\theta_1
\end{array}
\right]
=
\frac{1}{2}
\left[
\begin{array}{rr}
1&1\\
1&-1
\end{array}
\right]
\left[
\begin{array}{c}
\phi_0\\
\phi_1
\end{array}
\right]
\;.
\eeq

If we take the Hermitian
conjugate
of both sides of
Eq.(\ref{eq-decomp-1control}),
and then we replace
the angles $\phi_{b}$
and $\theta_{b}$
by their negatives,
we get

\beq
\exp\left(i\sum_{b\in Bool}
\phi_b\sigy\otimes P_b\right)=
\sigx(1)^{n(0)}
e^{i\theta_1\sigy(1)}
\sigx(1)^{n(0)}
e^{i\theta_0\sigy(1)}
\;.
\label{eq-decomp-1control-rev}
\eeq
Eq.(\ref{eq-decomp-1control-rev})
is equivalent to
decomposition (b) in
Fig.\ref{fig-1control}.
Thus,
decompositions (a)
and (b) in
Fig.\ref{fig-1control}
transform into each other under
time reversal.

\begin{figure}[h]
    \begin{center}
    \epsfig{file=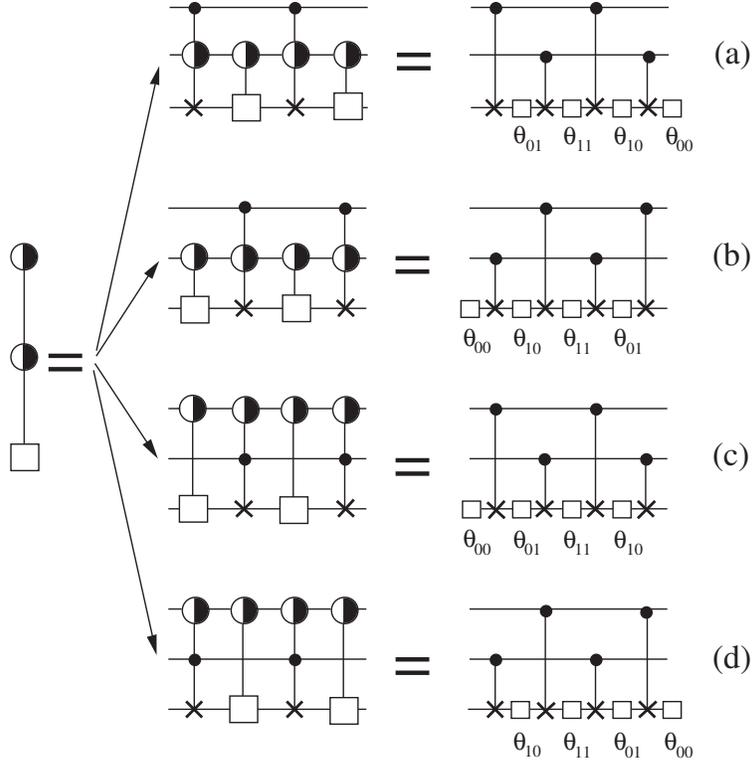, height=4.0in}
    \caption{
    Four possible decompositions
    of an $R_y(2)$-multiplexor
    with 2 controls.
    }
    \label{fig-2controls}
    \end{center}
\end{figure}

Fig.\ref{fig-2controls}
shows four possible ways of
decomposing an $R_y(2)$-multiplexor
with two controls.
Fig.\ref{fig-2controls}
was obtained
by applying the results of
Figs.\ref{fig-free-half-moon} and
\ref{fig-1control}.

In Fig.\ref{fig-2controls},
note that
decompositions (a) and (b)
transform into each other under
time reversal. Decompositions (c) and
(d) do too. Furthermore,
decompositions (b) and (c)
transform into each other under
bit permutation.

The decompositions
exhibited in
Fig.\ref{fig-2controls}
can also be expressed analytically.
For example,
decomposition (b)
is equivalent to:

\begin{eqnarray}
\lefteqn{\exp\left(i\sum_{b\in Bool^2}
\phi_{b}\sigy\otimes
P_{b}\right) =
}\nonumber\\
&&e^{i\theta_{00}\sigy(2)}
\sigx(2)^{n(1)}
e^{i\theta_{10}\sigy(2)}
\sigx(2)^{n(0)}
e^{i\theta_{11}\sigy(2)}
\sigx(2)^{n(1)}
e^{i\theta_{01}\sigy(2)}
\sigx(2)^{n(0)}
\;.
\label{eq-decomp-2-controls}
\end{eqnarray}
Eq.(\ref{eq-decomp-2-controls})
can be proven
using the same techniques
that were employed
in Eqs.(\ref{eq-decomp-1control-proof})
to prove Eq.(\ref{eq-decomp-1control}).
The proof requires that we assume:

\beq
\left[
\begin{array}{c}
\theta_{00}\\
\theta_{01}\\
\theta_{10}\\
\theta_{11}
\end{array}
\right]
=
\frac{1}{4}
\left[
\begin{array}{rrrr}
1&1&1&1\\
1&-1&1&-1\\
1&1&-1&-1\\
1&-1&-1&1
\end{array}
\right]
\left[
\begin{array}{c}
\phi_{00}\\
\phi_{01}\\
\phi_{10}\\
\phi_{11}
\end{array}
\right]
\;.
\eeq

\begin{figure}[h]
    \begin{center}
    \epsfig{file=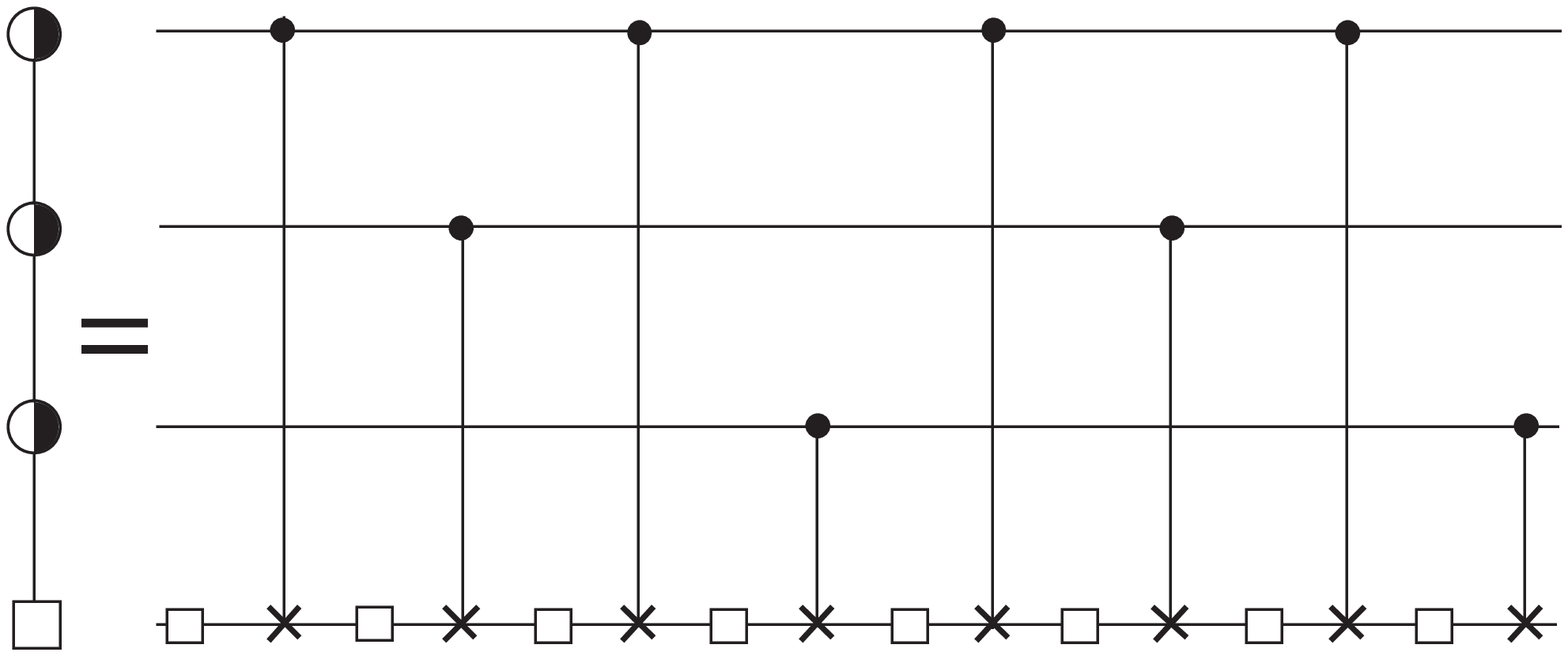, height=1in}
    \caption{
    One of several possible decompositions
    of an $R_y(2)$-multiplexor
    with 3 controls.
    }
    \label{fig-3controls}
    \end{center}
\end{figure}

\begin{figure}[h]
    \begin{center}
    \epsfig{file=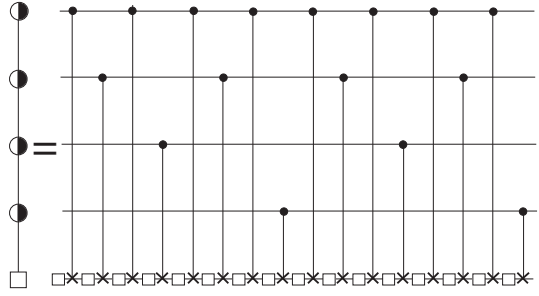, height=1.5in}
    \caption{
    One of several possible decompositions
    of an $R_y(2)$-multiplexor
    with 4 controls.
    }
    \label{fig-4controls}
    \end{center}
\end{figure}

Fig. \ref{fig-3controls}
(ditto, \ref{fig-4controls})
shows one of several possible decompositions
of an $R_y(2)$-multiplexor
 with 3 (ditto, 4) controls.
In general, decompositions for multiplexors with
$N_K$ controls can be obtained
starting from
decompositions for multiplexors
with $N_K-1$ controls.

\section{Approximation of Multiplexors}

In this section, we finally
define our approximation of
multiplexors. We give the
number of CNOTs
required to express the approximant,
and an upper
bound to the error incurred by
using it.

So far we have used $\nb$ to denote
a number of bits, and $\ns=2^\nb$
to denote the corresponding number of states.
Below, we will use two other numbers of bits,
$\nub$ and $\nub'$,
where
$\nub=\nb-1$ and $\nub'\leq \nub$.
Their corresponding numbers
of states will be denoted by
$\nus = 2^\nub$ and
$\nus' = 2^{\nub'}$.

Define an $\nus$ dimensional
 matrix $V$ by
\beq
V = H \pi_B \pi_G
\;,
\label{eq-v-fun-h}
\eeq
where $\pi_B$ is an
arbitrary bit permutation
on $\nub$ bits.
Eq.(\ref{eq-v-fun-h})
is a generalization of
Eq.(\ref{eq-w-fun-h}).
Both equations define
a new matrix (either $V$ or
the Walsh matrix $W$)
by permuting the columns of
the Hadamard matrix $H$.
$V$ becomes $W$
if we specialize
the bit permutation $\pi_B$ to
$\pi_R$.
If we denote the
columns of $V$
by $\vec{v}_j$
for $j\in \ZZ_{0, \nus-1}$,
then

\beq
\vec{v}_j=\vec{h}_{\pi_B\circ \pi_G(j)}
\;,
\label{eq-vi-def}
\eeq
which is the counterpart of
Eq.(\ref{eq-wi-def}).

In Ref.\cite{Tuc99},
the decomposition
of an $R_y(2)$
multiplexor
starts by taking
the following
Hadamard transform:

\beq
\vec{\theta} =
\frac{1}{\sqrt{\nus}}
H_{\nub}\vec{\phi}
\;,
\label{eq-theta-fun-phi}
\eeq
where $\nub=N_B-1$ and $\nus=2^\nub$.
The vectors $\{\vec{v}_i\}_{\forall i}$
constitute an orthonormal
basis for the space
$\RR^{\nus}$
in which $\vec{\phi}$ lives,
so $\vec{\phi}$
can always be expanded in terms
of them:

\beq
\vec{\phi}=
\sum_{i=0}^{\nus-1}
 \vec{v}_i
 ({\vec{v}_i}^\dagger \vec{\phi})
\;.
\eeq
Now suppose that we
truncate this expansion,
keeping only
the first $\nus'$ terms,
where $\nus'=2^{\nub'}$
and $\nub'\in \ZZ_{0,\nb-1}$.
Let us call $\vec{\phi'}$
the resulting approximation to
$\vec{\phi}$:

\beq
\vec{\phi'}=
\sum_{i=0}^{\nus'-1}
 \vec{v}_i
 ({\vec{v}_i}^\dagger \vec{\phi})
\;.
\label{eq-def-phi-prime}
\eeq
Define
$\vec{\theta'}$,
an approximation to
$\vec{\theta}$,
as follows:

\beq
\vec{\theta'} =
\frac{1}{\sqrt{\nus}}
H_{\nub}\vec{\phi'}
\;.
\label{eq-def-theta-prime}
\eeq
If we let $\{\vec{e}_i\}_{\forall i}$ denote
the standard basis vectors, then

\beq
H_{\nub}\vec{v}_i=
\left[
\begin{array}{c}
\vec{h}_0^\dagger\\
\vec{h}_1^\dagger\\
\vdots
\end{array}
\right]
\vec{h}_{\pi_B\circ \pi_G(i)}
=
\vec{e}_{\pi_B\circ \pi_G(i)}
\;.
\eeq
Therefore,

\beq
\vec{\theta'}=
\frac{1}{\sqrt{\nus}}
\sum_{i=0}^{\nus'-1}
\vec{e}_{\pi_B\circ \pi_G(i)}
(
{\vec{v}_i}^\dagger
\vec{\phi}
)
\;.
\label{eq-theta-prime}
\eeq

By virtue of Eq.(\ref{eq-theta-prime}),
if we list
the components
$\{\theta'_{b}\}_{\forall b}$
of $\vec{\theta'}$
in
the Grayish code order
specified
by the map
$\pi_B\circ \pi_G$,
then the items in the list
at positions from
$\nus'$  to the end of the list
are  zero.
Consider, for example,
Fig.\ref{fig-2controls},
which gives
the exact decompositions for
a multiplexor with
2 controls. Suppose
that in one of those
decompositions,
the angles $\theta_{b}$'s
in the second half (i.e., the half that does not
contain $\theta_{00}$)
of the decomposition
are
all zero.
Then the
one-qubit
rotations in the second half
of the decomposition
become the identity.
Then the three CNOTs in the second half
of the decomposition cancel each other
in pairs except for one CNOT that
survives. The net effect
is that the
decomposition for a multiplexor
with 2 controls
degenerates
into a decomposition for
a multiplexor
 with only 1 control.
The number of control bits is
reduced by one in this example.
In general,
we can approximate
a $U(2)$-multiplexor
by another $U(2)$-multiplexor
(the ``approximant")
that has fewer controls, and,
therefore, is expressible with fewer CNOTs.
We will
call the
 reduction in the number
of control bits the
{\bf bit deficit} $\delta_B$.
Hence, $\delta_B=\nub-\nub'$.

If $N_{CNOT}$ denotes the number of
CNOTs in an approximant with bit deficit
$\delta_B$, then it is clear from
Figs.\ref{fig-1control},
\ref{fig-2controls},
\ref{fig-3controls}
and
\ref{fig-4controls} that:

\beq
\begin{array}{|l|l|l|l|l|l|l|}
\hline
\delta_B=&0&1&2&\cdots&\nb-2&\nb-1\\
\hline
N_{CNOT}=&
2^{\nb-1}&
2^{\nb-2}&
2^{\nb-3}&
\cdots&
2&
0\\
\hline
\end{array}
\;.
\label{eq-n-cnots}
\eeq
Hence, for $\delta_B\in \mathbb{Z}_{0, \nb-2}$,
$N_{CNOT}= 2^{\nb-1-\delta_B}$,
but for
$\delta_B=\nb-1$,
$N_{CNOT}= 0$.

The bit permutation
$\pi_B$
on which the approximation of
a multiplexor depends
can be chosen
according to various
criteria.
If we choose $\pi_B=\pi_R$,
then our approximation
will keep only the
higher constancy components
of $\vec{\phi}$.
Such a smoothing,
{\bf high constancies approximation}
might be useful for some tasks.
Similarly, if we choose $\pi_B=1$,
then our approximation will
keep only the lower constancy
components of $\vec{\phi}$,
giving a {\bf low constancies approximation}.
Alternatively, we could use for
$\pi_B$ a bit
permutation,
out of all possible
bit permutations on $\nub$ bits,
that minimizes
the distance between the
original multiplexor and its
approximant.
Such a {\bf dominant constancies
approximation} is useful if
our goal is to minimize
the error incurred by the approximation.

The error incurred by
approximating a multiplexor
can be bounded above as follows.
Let
$\{e^{i\phi_b\sigy}\}_{\forall b\in Bool^{\nub}}$
denote
the $R_y(2)$-subset of an
$R_y(2)$-multiplexor $\Upsilon$
and
$\{e^{i\phi_b'\sigy}\}_{\forall b\in Bool^{\nub}}$
that of its approximant $\Upsilon'$.
Call $\|\Upsilon'-\Upsilon\|_2$
 the error of approximating
$\Upsilon$ by $\Upsilon'$. Note that

\begin{subequations}
\label{eq-error}
\begin{eqnarray}
\|\Upsilon'-\Upsilon\|_2&=&
\|\oplus_{b\in Bool^{\nub}} (e^{i\phi'_b\sigy}
-e^{i\phi_b\sigy})\|_2\\
&=&\max_b
\|e^{i\phi'_b\sigy}
-e^{i\phi_b\sigy}\|_2\\
&\leq&
\max_b |\phi'_b-\phi_b|=\|\vec{\phi'}-\vec{\phi}\|_\infty
\;.
\label{eq-step-ineq}
\end{eqnarray}
\end{subequations}
To arrive at step Eq.(\ref{eq-step-ineq}),
we used the results of Appendix \ref{ap-ineq}.
We will sometimes refer to
$\|\vec{\phi'}-\vec{\phi}\|_\infty$
as the linearized error, to distinguish it
from the error $\|\Upsilon'-\Upsilon\|_2$.

A simple picture emerges from all this.
The error $\epsilon$
 and the number of CNOTs $N_{CNOT}$
 are two  costs that we would
 like to minimize. These two costs are
 fungible to a certain extent.
Given a multiplexor $\Upsilon$,
and an upper bound
$\epsilon_0$ on $\epsilon$,
we can use
Eqs.(\ref{eq-n-cnots}) and (\ref{eq-error})
to find the approximant $\Upsilon'$
with the smallest $N_{CNOT}$.
Similarly,
given a multiplexor $\Upsilon$,
and an upper bound
$(N_{CNOT})_0$ on $N_{CNOT}$,
we can use
Eqs.(\ref{eq-n-cnots}) and (\ref{eq-error})
to find the approximant $\Upsilon'$
with the smallest $\epsilon$.

At this point we encourage
the reader to read Appendix
\ref{ap-computer-results}. It
discusses the output of a computer
program
that
calculates $\vec{\phi'}$
from $\vec{\phi}$
via
Eq.(\ref{eq-def-phi-prime}).

Next we will show that
 Eq.(\ref{eq-def-phi-prime})
can be simplified considerably
by taking into account the explicit values of
the column vectors $\vec{v}_j$.

To get a quick glimpse of
the  simplification
we seek,
consider first the special case $\nub=2$.
We have

\beq
H_2 =
\frac{1}{2}
\begin{array}{|rrrr|}
{\scriptstyle\vec{h}_{00}}&
{\scriptstyle\vec{h}_{01}}&
{\scriptstyle\vec{h}_{10}}&
{\scriptstyle\vec{h}_{11}}
\\
\hline
\pone&\pone&\pone&\pone
\\
\pone&\none&\pone&\none
\\
\pone&\pone&\none&\none
\\
\pone&\none&\none&\pone
\\
\hline
\end{array}
\;,\;\;
W_2 =
\frac{1}{2}
\begin{array}{|rrrr|}
{\scriptstyle\vec{h}_{00}}&
{\scriptstyle\vec{h}_{10}}&
{\scriptstyle\vec{h}_{11}}&
{\scriptstyle\vec{h}_{01}}
\\
\hline
\pone&\pone&\pone&\pone
\\
\pone&\pone&\none&\none
\\
\pone&\none&\none&\pone
\\
\pone&\none&\pone&\none
\\
\hline
\end{array}
\;.
\eeq
Define a matrix $\mu$ by

\beq
\mu =
\left(
\begin{array}{cc}
1 & 1\\
1 & 1
\end{array}
\right)
\;.
\eeq
For any matrix $A$,
let $A(:,i:j)$ be the submatrix
of $A$ obtained
by keeping only
its columns from $i$ to $j$.
It is easy to check that

\beq
W_2(:,0:3) W_2(:,0:3)^T = diag(1,1,1,1)
\;,
\eeq

\beq
W_2(:,0:1) W_2(:,0:1)^T
=
\frac{1}{2}
\left(
\begin{array}{cc}
\mu & 0 \\
0 & \mu
\end{array}
\right)
=
\frac{1}{2}
(\mu \oplus \mu)
\;,
\label{eq-w2-two-cols}
\eeq
and

\beq
W_2(:,0:0) W_2(:,0:0)^T
=
\frac{1}{4}
\left(
\begin{array}{cc}
\mu & \mu \\
\mu & \mu
\end{array}
\right)
=
\frac{1}{4}
(\mu \otimes \mu)
\;.
\eeq
In each case, we formed a
``decimated matrix" $W(:, 0:\nus'-1)$ from $W$,
where $\nus'=2^{\nub'}$.
Then we showed that the projection operator
$W(:, 0:\nus'-1)W(:, 0:\nus'-1)^T$
onto the column space of the decimated
matrix,
is a matrix whose entries are all
either 0 or
$\frac{1}{2^{\delta_B}}$,
and these entries
sum to one along each row (or column).
Given a set $S$ of real numbers,
and given $S_1\subset S$,
call the average of the elements of $S_1$
a ``partial average" of the elements of $S$.
For example, if $\nus'=2$
and $\pi_B=\pi_R$, then $V=W$ and
$\vec{\phi'} = W_2(:,0:1)W_2(:,0:1)^T\vec{\phi}$.
From Eq.(\ref{eq-w2-two-cols}),
the entries of $\vec{\phi'}$
are partial averages of
the entries of $\vec{\phi}$.

Next we show how to simplify
Eq.(\ref{eq-def-phi-prime})
for arbitrary $\nub$, not just
for $\nub=2$.

Define an $\nus$ dimensional
matrix $\Gamma^{(\nub, \nub')}$ by

\beq
\Gamma^{(\nub, \nub')}_{qs}=
\sum_{r=0}^{\nus'-1}
(W_\nub)_{qr}
(W_\nub^T)_{rs}
\;.
\label{eq-gamma-fun-w}
\eeq
Below, we will show that
$\Gamma^{(\nub, \nub')}$
reduces to:

\beq
\Gamma^{(\nub, \nub')}_{qs}=
\frac{1}{2^{\delta_B}}
\theta(
\floor{
\frac{q}{2^{\delta_B}}
}
=
\floor{
\frac{s}{2^{\delta_B}}
}
)
\;.
\label{eq-gamma-fun-floor}
\eeq
For example, when $\nub=2, \nub'=1$,
Eq.(\ref{eq-gamma-fun-floor}) becomes

\beq
\left[\Gamma^{(2,1)}_{qs}\right]
=
\frac{1}{2}
\left[\theta(
\floor{\frac{q}{2}}=
\floor{\frac{s}{2}}
)\right]
=
\frac{1}{2}
\begin{array}{c|cccc}
&{\scriptscriptstyle 00}&{\scriptscriptstyle 01}
&{\scriptscriptstyle 10}&{\scriptscriptstyle 11}\\
\hline
{\scriptscriptstyle 00}&1&1&0&0\\
{\scriptscriptstyle 01}&1&1&0&0\\
{\scriptscriptstyle 10}&0&0&1&1\\
{\scriptscriptstyle 11}&0&0&1&1
\end{array}
\;.
\eeq

Eqs.(\ref{eq-gamma-fun-w})
and (\ref{eq-gamma-fun-floor})
are given in component form.
The identical statements
written in matrix form are,
respectively,

\beqa
\Gamma^{(\nub, \nub')}
&=&
W_\nub(:,0:\nus'-1)
W_\nub(:,0:\nus'-1)^T\\
&=&
W_\nub
\left[
\begin{array}{cc}
I_{\nus'}& 0\\
0 & 0_{\nus-\nus'}
\end{array}
\right]
W_\nub^T
\;,
\eeqa
and

\beq
\Gamma^{(\nub, \nub')}
=
\frac{1}{2^{\delta_B}}
(\mu^{\otimes \delta_B})^{\oplus\nus'}
\;.
\eeq

Eq.(\ref{eq-gamma-fun-floor})
can be proven from
Eq.(\ref{eq-gamma-fun-w}) as follows:

\beqa
\Gamma^{(\nub, \nub')}_{qs}&=&
\frac{1}{\nus}
\sum_{r=0}^{\nus'-1}
(-1)^{
\sum_{\alpha=0}^{\nub-1}
\sum_{\beta=0}^{\nub-1}
(q_\alpha r_\beta + r_\beta s_\alpha)
(
\delta_{\alpha + \beta}^{ \nub-1}+
\delta_{\alpha + \beta}^{ \nub}
)
}
\\ &=&
\frac{1}{\nus}
\left(
\sum_{r_{\nub-1}=0}^{1}
\cdots
\sum_{r_1=0}^{1}
\sum_{r_0=0}^{1}
\right)
\theta(
r_{\nub-1} = r_{\nub-2} =\cdots
=r_{\nub-\delta_B}=0)
\nonumber\\&&
(-1)^{
\sum_{\alpha=0}^{\nub-1}
\sum_{\beta=0}^{\nub-1}
r_\beta t_\alpha
(
\delta_{\alpha + \beta}^{ \nub-1}+
\delta_{\alpha + \beta}^{ \nub}
)
}\;\;({\rm where}\;\; t = q\oplus s)
\\ &=&
\frac{1}{\nus}
\left(
\sum_{r_{\nub-1}=0}^{1}
\cdots
\sum_{r_1=0}^{1}
\sum_{r_0=0}^{1}
\right)
\theta(
r_{\nub-1} = r_{\nub-2} =\cdots
=r_{\nub-\delta_B}=0)
\nonumber\\&&
(-1)^{
r_{\nub-1}(t_0  + t_1) +
r_{\nub-2}(t_1  + t_2) +
\cdots
r_1(t_{\nub-2}  + t_{\nub-1}) +
r_0(t_{\nub-1})
}
\\
&=&
\frac{1}{\nus} 2^{\nub-\delta_B}
\delta_{q_{\nub-1}}^{s_{\nub-1}}
\delta_{q_{\nub-2}}^{s_{\nub-2}}
\cdots
\delta_{q_{\delta_B}}^{s_{\delta_B}}
\label{eq-delta-fun-step}
\\
&=&
\frac{1}{2^{\delta_B}}
\theta(
\floor{
\frac{q}{2^{\delta_B}}
}
=
\floor{
\frac{s}{2^{\delta_B}}
}
)
\;.
\eeqa
To arrive at Eq.(\ref{eq-delta-fun-step}),
we used Eq.(\ref{eq-bool-delta-fun}).

Recall that $W$ and $V$
can both we obtained by permuting
the columns of $H$:

\beq
W = H \pi_R \pi_G
\;,
\eeq
and

\beq
V = H \pi_B \pi_G
\;.
\eeq
From these two equations and
from the fact, proven earlier, that
$\pi_B$ and $\pi_R$ commute with $H$, we get

\beq
V = \pi_B \pi_R H \pi_R \pi_G = \pi_B \pi_R W
\;.
\eeq
Thus,

\beqa
\vec{\phi'}
&=&
V_\nub
\left[
\begin{array}{cc}
I_{\nus'}& 0\\
0 & 0_{\nus-\nus'}
\end{array}
\right]
V_\nub^T
\vec{\phi}\\
&=&
\pi_B \pi_R
W_\nub
\left[
\begin{array}{cc}
I_{\nus'}& 0\\
0 & 0_{\nus-\nus'}
\end{array}
\right]
W_\nub^T
\pi_R \pi_B^T \vec{\phi}
\;.
\eeqa
Hence, if we define
$\vec{\psi}$ and $\vec{\psi'}$ by

\beq
\vec{\psi} = \pi_R \pi_B^T \vec{\phi}
\;,\;\;
\vec{\psi'} = \pi_R \pi_B^T \vec{\phi'}
\;,
\eeq
then

\beq
\vec{\psi'} = \Gamma^{(\nub, \nub')}\vec{\psi}
\;.
\label{eq-Phi-pr-fun-Phi}
\eeq

We see
from Eq.(\ref{eq-Phi-pr-fun-Phi})
that even when $\pi_B\neq \pi_R$,
the entries of
$\vec{\psi'}$
(which are the same as the entries of
$\vec{\phi'}$ but in a different
order)
are partial averages of the entries of
$\vec{\psi}$ (which are the same as
those of $\vec{\phi}$
but in a different
order). Thus

\beq
\min_k(\phi_k) \leq \phi_j' \leq \max_k (\phi_k)
\;,
\eeq
for all $j\in \ZZ_{0, \nus-1}$.
This last equation
implies

\beq
|\phi'_j - \phi_j |
\leq \max_k(\phi_k) - \min_k(\phi_k)
\;,
\eeq
for all $j$.

As we mentioned before,
the quantum compiling algorithm
of Ref.\cite{Tuc04Oct}
uses $U(2)$-multiplexors
that are more general than
 the $R_y(2)$-multiplexors
considered above.
Luckily, the above
results for $R_y(2)$-multiplexors
are still valid, with minor
modifications, for the
more general ones.
Indeed, the $U(2)$-subset of
the multiplexors used
in Ref.\cite{Tuc04Oct}
is of the form
$\{ \exp[i(\phi_{1b}\sigma_{s_1}
+ \phi_{2b}\sigma_{s_2})]
\sigma_w^{f(b)}\}_{\forall b}$.
Ref.\cite{Tuc04Oct}
defines vectors
$\vec{\phi}_1$
and
$\vec{\phi}_2$
from the parameters
$\{\phi_{1b}\}_{\forall b}$
and
$\{\phi_{2b}\}_{\forall b}$,
respectively.
It then defines
$\vec{\theta_1}$
and
$\vec{\theta_2}$
as Hadamard transforms
of
$\vec{\phi_1}$
and
$\vec{\phi_2}$,
respectively,
just as Eq.(\ref{eq-theta-fun-phi})
defines $\vec{\theta}$ as a
Hadamard transform of $\vec{\phi}$.
We can define approximations
$\vec{\phi'_1}$ and
$\vec{\theta'_1}$
by replacing
$\vec{\phi}$,
$\vec{\theta}$,
$\vec{\phi'}$,
$\vec{\theta'}$
by
$\vec{\phi_1}$,
$\vec{\theta_1}$,
$\vec{\phi'_1}$,
$\vec{\theta'_1}$,
respectively,
within
Eqs.(\ref{eq-def-phi-prime})
and (\ref{eq-def-theta-prime}).
We can define approximations
$\vec{\phi'_2}$ and
$\vec{\theta'_2}$
analogously.
The expansions of
$\vec{\phi'}_1$
and
$\vec{\phi'}_2$
in the $\vec{v}_i$
basis can be truncated
at the same $\nus'$.
The table given in
Eq.(\ref{eq-n-cnots})
for the number of CNOTs
still applies, except that
$N_{CNOT}$ may change by 1
if we eliminate the
$\theta_{0\ldots 00}$ gate
as in Ref.\cite{Tuc04Oct}.
When a $U(2)$-multiplexor $\Upsilon$
with $U(2)$-subset
$\{ \exp[i(\phi_{1b}\sigma_{s_1}
+ \phi_{2b}\sigma_{s_2})]
\sigma_w^{f(b)}\}_{\forall b}$
is approximated by a $U(2)$-multiplexor
$\Upsilon'$ with $U(2)$-subset
$\{ \exp[i(\phi'_{1b}\sigma_{s_1}
+ \phi'_{2b}\sigma_{s_2})]
\sigma_w^{f(b)}\}_{\forall b}$,
one can show, using the results
of Appendix \ref{ap-ineq}, that

\beq
\|\Upsilon'-\Upsilon\|_2
\leq
\max_b \sqrt{\sum_{j=1}^2(\phi'_{jb}-\phi_{jb})^2}
\;,
\eeq
which is a generalization of Eq.(\ref{eq-error}).

\appendix
\section{Appendix: Distance between \\
two $SU(2)$ matrices}
\label{ap-ineq}

In this appendix, we
establish a well known(see Ref.\cite{Golub},
page 574)
 upper bound for
the distance
(measured in either the 2-norm or the
Frobenius norm)
between two $SU(2)$ matrices.

Let
$\vec{\alpha}, \vec{\alpha'}$
be 3d real vectors.
Define
$
\vec{\Delta \alpha}
 = \vec{\alpha'}-\vec{\alpha}
$. If $|\vec{\Delta \alpha}|<<1$,
then

\beqa
\|e^{i\vec{\alpha'}\cdot \vsig}
-e^{i\vec{\alpha}\cdot \vsig}\|_2&=&
\|e^{i\vec{\alpha'}\cdot \vsig}
e^{-i\vec{\alpha}\cdot \vsig}
-1\|_2
\\ &\approx &
\|i\vec{\Delta\alpha}\cdot \vsig\|_2
\\&=&
|\vec{\Delta\alpha}|
\;.
\eeqa
Next, we will show that
this approximation can be turned into an
inequality.

Consider first  the
 special case where $\vec{\alpha}$
and $\vec{\alpha'}$ both point in the
Y direction.  Then

\beqa
\|e^{i\alpha'\sigy} - e^{i\alpha\sigy}\|_2
&=&
\|e^{i\Delta\alpha\sigy}-1\|_2\\
&=&
\|
\left(
\begin{array}{cc}
\cos(\Delta \alpha) -1 & \sin(\Delta\alpha) \\
-\sin(\Delta\alpha) & \cos(\Delta \alpha) -1
\end{array}
\right)
\|_2
\\
&=&
\|
2\sin(\frac{\Delta \alpha}{2})
\left(
\begin{array}{cc}
-\sin(\frac{\Delta \alpha}{2})
& \cos(\frac{\Delta \alpha}{2}) \\
-\cos(\frac{\Delta \alpha}{2})
& -\sin(\frac{\Delta \alpha}{2})
\end{array}
\right)\|_2
\\
&=&2 |\sin(\frac{\Delta\alpha}{2})|\\
&\leq& |\Delta \alpha|
\;.
\eeqa

To find an upper bound for $\|e^{i\vec{\alpha'}\cdot \vsig}
-e^{i\vec{\alpha}\cdot \vsig}\|_2$ when
either $\vec{\alpha}$
or $\vec{\alpha'}$ does not point in the
Y direction, we will use the following identity.
For $A, E\in \CC^{n\times n}$
and $t\in \RR^{\geq 0}$,

\beq
e^{(A+E)t} - e^{At}
=
\int_0^t ds\;
e^{A(t-s)}
E
e^{(A+E)s}
\;.
\label{eq-diff-eq-sol}
\eeq
To prove Eq.(\ref{eq-diff-eq-sol}),
let ${\cal L}$ and ${\cal R}$
stand for the left and right hand sides of
Eq.(\ref{eq-diff-eq-sol}). It is easy to
verify that

\beq
({\cal L - R})(0)=0
\;,\;\;
\frac{d({\cal L - R})}{dt}=A({\cal L - R})
\;.
\eeq
This initial value problem
has the unique solution ${\cal L-\cal R}=0$.

In Eq.(\ref{eq-diff-eq-sol}),
set  $A = i\vec{\alpha}\cdot\vsig$,
and $E = i\vec{\Delta \alpha}\cdot\vsig$,
and take the 2-norm of both sides.
This yields

\beqa
\|
e^{i(\vec{\alpha} +
\vec{\Delta\alpha})
\cdot \vsig}
-
e^{i\vec{\alpha}
\cdot \vsig}
\|_2
&\leq&
\int_0^1
ds\;
\|
e^{i \vec{\alpha}
\cdot \vsig(1-s)}\|_2
\|
i
\vec{\Delta\alpha}
\cdot \vsig
\|_2
\|
e^{i(\vec{\alpha} +
\vec{\Delta\alpha})
\cdot \vsig s}
\|_2
\\
&=&
|\vec{\Delta\alpha}|
\;.
\eeqa

One can also find an
upper bound for the
distance, in the
Frobenius norm, between
 two $SU(2)$ matrices.
If $U=e^{i\vec{\alpha}\cdot \vsig}$
and $U'=e^{i\vec{\alpha'}\cdot \vsig}$,
then the eigenvalues of $U$  are
$e^{i\theta}, e^{-i\theta}$,
for some real number $\theta$.
Likewise, the eigenvalues of
$U'$  are
$e^{i\theta'}, e^{-i\theta'}$.
Thus $\tr(U'-U)$ is real. If
we denote the eigenvalues of $U'-U$
by $x\pm i y $ with $x,y\in \RR$,
then $(U'-U)^\dagger (U'-U)$ has a single eigenvalue
 $x^2+y^2$ with algebraic multiplicity 2.
Thus

\beq
\|e^{i\vec{\alpha'}\cdot \vsig}
-e^{i\vec{\alpha}\cdot \vsig}\|_F
=\sqrt{2}\;
\|e^{i\vec{\alpha'}\cdot \vsig}
-e^{i\vec{\alpha}\cdot \vsig}\|_2
\;.
\label{eq-2-f-norms}
\eeq
But we've already proven that
$\|e^{i\vec{\alpha'}\cdot \vsig}
-e^{i\vec{\alpha}\cdot \vsig}\|_2$
is bounded above
by $|\vec{\Delta \alpha}|$ so

\beq
\|e^{i\vec{\alpha'}\cdot \vsig}
-e^{i\vec{\alpha}\cdot \vsig}\|_F
\leq
\sqrt{2}\;|\vec{\Delta \alpha} |
\;.
\eeq

\section{Appendix: Computer Results}
\label{ap-computer-results}

In this appendix, we discuss
a simple computer program
that verifies and illustrates many of the
results of this paper.
Our program is written in the Octave
language.
Octave is a gratis, open-source
interpreter
that understands a subset of the Matlab
language. Hence,
our program should also run in a Matlab
environment with few or no modifications.

Our main m-file is called \verb=my_moo.m=.
When you run \verb=my_moo=, Octave
produces two output files called
\verb=out_phis.txt= and \verb=out_error.txt=.

A typical \verb=out_phis.txt=
file reads:
{\scriptsize
\begin{verbatim}
phi(1)= 0.133765891
phi(2)= 0.270447403
phi(3)= 0.307625920
phi(4)= 0.311291575
phi(5)= 0.452735037
phi(6)= 0.569045961
phi(7)= 0.653136015
phi(8)= 0.867156088
-----------------------
permutation 1 = (1,2,3)
delta_B, phi_prime=
0   0.134   0.270   0.308   0.311   0.453   0.569   0.653   0.867
1   0.293   0.420   0.480   0.589   0.293   0.420   0.480   0.589
2   0.387   0.504   0.387   0.504   0.387   0.504   0.387   0.504
3   0.446   0.446   0.446   0.446   0.446   0.446   0.446   0.446
-----------------------
permutation 2 = (1,3,2)
delta_B, phi_prime=
0   0.134   0.270   0.308   0.311   0.453   0.569   0.653   0.867
1   0.293   0.420   0.480   0.589   0.293   0.420   0.480   0.589
2   0.356   0.356   0.535   0.535   0.356   0.356   0.535   0.535
3   0.446   0.446   0.446   0.446   0.446   0.446   0.446   0.446
-----------------------
permutation 3 = (2,1,3)
delta_B, phi_prime=
0   0.134   0.270   0.308   0.311   0.453   0.569   0.653   0.867
1   0.221   0.291   0.221   0.291   0.553   0.718   0.553   0.718
2   0.387   0.504   0.387   0.504   0.387   0.504   0.387   0.504
3   0.446   0.446   0.446   0.446   0.446   0.446   0.446   0.446
-----------------------
permutation 4 = (2,3,1)
delta_B, phi_prime=
0   0.134   0.270   0.308   0.311   0.453   0.569   0.653   0.867
1   0.202   0.202   0.309   0.309   0.511   0.511   0.760   0.760
2   0.356   0.356   0.535   0.535   0.356   0.356   0.535   0.535
3   0.446   0.446   0.446   0.446   0.446   0.446   0.446   0.446
-----------------------
permutation 5 = (3,1,2)
delta_B, phi_prime=
0   0.134   0.270   0.308   0.311   0.453   0.569   0.653   0.867
1   0.221   0.291   0.221   0.291   0.553   0.718   0.553   0.718
2   0.256   0.256   0.256   0.256   0.636   0.636   0.636   0.636
3   0.446   0.446   0.446   0.446   0.446   0.446   0.446   0.446
-----------------------
permutation 6 = (3,2,1)
delta_B, phi_prime=
0   0.134   0.270   0.308   0.311   0.453   0.569   0.653   0.867
1   0.202   0.202   0.309   0.309   0.511   0.511   0.760   0.760
2   0.256   0.256   0.256   0.256   0.636   0.636   0.636   0.636
3   0.446   0.446   0.446   0.446   0.446   0.446   0.446   0.446
\end{verbatim}
}

The corresponding \verb=out_error.txt=
file
reads
{\scriptsize
\begin{verbatim}
error as function of (permutation\delta_B)
            0           1           2           3
   1    1.110e-16   2.779e-01   3.627e-01   4.215e-01
   2    1.110e-16   2.779e-01   3.324e-01   4.215e-01
   3    1.110e-16   1.491e-01   3.627e-01   4.215e-01
   4    1.110e-16   1.070e-01   3.324e-01   4.215e-01
   5    5.551e-17   1.491e-01   2.316e-01   4.215e-01
   6    5.551e-17   1.070e-01   2.316e-01   4.215e-01
\end{verbatim}
}

In this example, $\nb=4$
so $\nub=3$ and $\nus=8$.
The first 8 lines of
\verb=out_phis.txt= give the
components of $\vec{\phi}$.
In this case,
the computer picked 8 independent random
numbers from the unit interval,
and then it sorted them in
non-decreasing order.
\verb=my_moo.m= can be easily modified
so as to allow the user himself to
supply the components of $\vec{\phi}$.

After listing $\vec{\phi}$,
\verb=out_phis.txt=
lists the $\nub!$
permutations $\pi_B$ of  $\nub$ bits.
For each $\pi_B$, it prints
the components of $\vec{\phi'}$,
listed as a row, for each
value of
 $\delta_B$(=row label).
Note that for $\delta_B=0$,
$\vec{\phi'}=\vec{\phi}$,
and for $\delta_B=\nub$,
all $\phi'_j$ are equal to the average
of all the components of $\vec{\phi}$.
Note also that for all values of
$\delta_B$ and $j$,
$\phi_j'\in [\min_k(\phi_k), \max_k(\phi_k)]$.

The second output file, \verb=out_error.txt=,
gives a table of the
linearized
error $\|\vec{\phi'}-\vec{\phi}\|_\infty$
as a function
of permutation number(=row label)
and $\delta_B$(=column label).
As expected, the
error is zero
when $\delta_B$ is zero, and it is independent
of the permutation $\pi_B$
 when $\delta_B$ is maximum
(When the bit deficit
$\delta_B$ is maximum, the approximant
has no control bits, so
permuting bits at positions
$Z_{0, \nub-1}$ does not affect the error.)

Note
that in the above example,
the last permutation
minimizes the  error for all $\delta_B$.
This last permutation is
$\pi_B=\pi_R=$ (bit-reversal), and it
gives
a high constancies
expansion.
Recall that for this  example,
\verb=my_moo.m=
generated iid (independent, identically
distributed) numbers for
the components of $\vec{\phi}$,
and then it rearranged them in monotonic
order. When
$\vec{\phi}$
is chosen in this way,
the graph $\{(j, \phi_j)\}_{\forall j}$
has a high probability of lying close to a
straight line, and a high constancy
staircase is the best fit for
a straight line. For this reason,
almost every time that \verb=my_moo.m=
is operated in the mode which
generates iid numbers for
the components of $\vec{\phi}$,
the high constancies expansion
minimizes the error for all $\delta_B$.
However, this need not always occur,
as the following counterexample shows.
Try running \verb=my_moo.m=
for  $\nb=5$, and
for $\vec{\phi}$ with its first
7 components equal to 0 and its 9 subsequent
components equal to 1.
For this $\vec{\phi}$, and for $\delta_B=3$,
the high constancies
expansion yields an error of 7/8 while
some of the other expansions
yield errors as low as 5/8.

Note that although \verb=my_moo.m=
visits all $\nub!$ permutations
of the control bits,
visiting all permutations is
a very inefficient way of
finding the minimum error.
In fact,
the $\nub!$
control bit permutations can be
grouped into equivalence classes,
such that all permutations in a
 class give the same error.
It's clear from Fig.\ref{fig-approx}
that we only have to visit
$\binom{\nub}{  \delta_B} =
\frac{\nub!}{\delta_B!\nub'!}$
(recall $\nub' = \nub-\delta_B$)
equivalence classes of permutations.
Whereas $\nub!\approx\nub^\nub=e^{\nub \ln \nub}$
is exponential in $\nub$,
$\binom{\nub}{  \delta_B}$ is polynomial
in $\nub$ for two very important extremes. Namely,
when $\delta_B$ or
$\nub'$ is
of order one whereas $\nub$ is very large.
Indeed, if $\delta_B=1$ or
$\nub'=1$, then
$ \binom{\nub}{ \delta_B} = \nub$;
if $\delta_B=2$ or
$\nub'=2$, then
$\binom{\nub}{  \delta_B} =
\frac{\nub(\nub-1)}{1\cdot2}$, etc.

\end{document}